\documentclass[amstex,epsf,amssymb,usenatbib]{mn2e}
\usepackage{epsf}
\usepackage{amsmath,amssymb}
\def\aj{AJ}                   
\def\araa{ARA\&A}             
\def\apj{ApJ}                 
\def\apjl{ApJ}                
\def\apjs{ApJS}               
\def\aap{A\&A}                
\def\mnras{MNRAS}             

\title{Hydrodynamic simulations of molecular outflows driven by slow-precessing
protostellar jets \\}

\author[M.D. Smith \& A. Rosen]
{Michael D. Smith $^1$\thanks{E-mail: mds@arm.ac.uk}
\& {Alexander Rosen $^{2}$\thanks{E-mail: alex.rosen@dcu.ie}}\\
$^1$Armagh Observatory, College Hill, Armagh BT61 9DG, Northern Ireland\\
$^2$Dublin City Univerity, School of Mathematics, Glasnevin, Dublin 9\\
}

\date{Accepted .....
      Received ..... ;
      in original form .....}

\pagerange{\pageref{firstpage}--\pageref{lastpage}}
\pubyear{2004}

\begin{document}

\maketitle

\label{firstpage}

\begin{abstract}

We present hydrodynamic simulations of molecular outflows driven by jets with
a long period of precession, motivated by observations of arc-like features and
S-symmetry in outflows associated with young stars. We
simulate images of not only H$_2$ vibrational and CO rotational
emission lines, but also of atomic emission. The density cross section
displays a jaw-like cavity, independent of precession rate.
In molecular hydrogen, however,
we find ordered chains of bow shocks and meandering streamers which contrast
with the chaotic structure produced by jets in rapid
precession. A feature particularly dominant in atomic emission
is a stagnant point in the flow that remains near the inlet
and alters shape and brightness as the jet skims by.
Under the present conditions, slow jet precession yields a relatively
high fraction of mass accelerated to high speeds, as also attested to in
simulated CO line profiles. Many outflow 
structures, characterised by HH\,222 (continuous ribbon), HH\,240 (asymmetric
chains of bow shocks) and RNO\,43N (protruding cavities), are probably
related to the slow precession model.

\end{abstract}

\begin{keywords}
 hydrodynamics -- shock waves -- ISM: clouds -- ISM: jets and outflows
-- ISM: molecules
\end{keywords}

\section{Introduction}              

Collimated outflows are integrally connected with star formation 
\citep{2000prpl.conf..867R, 2001ARA&A..39..403R}. The outflowing mass is distributed 
over a large volume and can far exceed the suspected mass of the growing star.
This suggests that momentum is generated deep within the potential well
of the young star and that, after extraction, is transmitted into the ambient gas.
Jets of molecules are often associated with the youngest protostars. However,
one view suggests that these jets are highly ballistic and that the low
density of the surroundings does not efficiently soak up the jet momentum. As the
protostars age and the accretion rate declines, their  molecular jets become
less dense and more atomic which may increase the efficiency. Alternatively,
the jet thrust might be sprayed into a large volume through precession of the
direction of release. The consequences of such a scenario are explored here.

Observations of protostellar jets frequently provide evidence for precession.  
Some examples  with precession angles in the range
5\raisebox{0.9ex}{$\circ$} -- 20\raisebox{0.9ex}{$\circ$} are
Cep~A \citep{1999A&A...343..571G}; Cep~E \citep{1996AJ....112.2086E};
HH~31 \citep{1997AJ....114.1138G}; HH~34 \citep{1994ApJ...428L..65B}; and
HH~333 \citep{2001ApJ...554..132A}. Larger precession angles are suggested for 
L\,1228 \citep{1995ApJ...454..345B}, PV~Cep \citep{1997AJ....114..265Ga} and 
IRAS~20126+4104 \citep{2000ApJ...535..833S}.

We present here a sequence of three dimensional hydrodynamic simulations.
They were executed with a modified version of ZEUS-3D which solves a detailed
cooling function and a reduced chemical network (see \S\,\ref{method}).
We drive a molecular jet into a uniform molecular environment. Previously, we have
examined the physical and observational structure which results for jets with (i)
a range of densities relative to the ambient density \citep{rs1},
(ii) significant sustained changes to the inlet mass flux \citep{rs2} and
(iii) a range of precession angles but all with a fast precession period \citep{rs3}.
Fast jet precession, however, does not produce the S-symmetry as observed in the
above examples. Instead, the jet impact describes an expanding ring which fragments into a
multitude of shocks. In addition, a stagnation region appears once the impact no longer 
encounters the precession axis, as examined by \citet{2001MNRAS.327..507L} for an atomic model.
The ambient gas trapped in this region may lead to observable features.

Here, we investigate the consequences of jets undergoing slow precession, the precession speed being
defined in terms of the appropriate dynamical time (see \S \ref{method}). We wish to
identify the global structures associated with  a range of commonly observed
emission lines. Curved outflows of various types are indeed observed. The very young protostar VLA~1623
is associated with a slender tube-like CO outflow which meanders
through the $\rho$~Ophiuchi cloud \cite{1995MNRAS.277..193D}.
A few bipolar outflows show C-type or mirror symmetry rather than S-type or point symmetry
(e.g. the S233IR outflow, \citealt{2002A&A...387..931B}). Such flows could result from twin jets
associated with one protostar within a close binary: the space velocities of both jets
at each instant are the vector additions of  fixed components in the protostar's frame and
the variable orbital velocity \citep{1998ApJ...494L..79F, 2002ApJ...568..733M}.
There are even some theories that tie the precession of jets to the warping of the
disk \citep{2003ApJ...591L.119L}.  The star formation process, which must be able to
generate a large number of binaries, thus includes potential mechanisms for varying
the direction of ejection.

Besides identifying outflow properties, there are two issues we try to resolve.
Firstly, strong H$_2$ 1--0~S(1) emission is often measured in the near-infrared
at 2.12\,$\mu$m. Taking this to be ten percent of the total H$_2$ emission, which is 
taken to be ten per cent of the total radiation, then the jet power must be at least 
100 times stronger. This can just be reconciled with current theories for young, 
Class~0 protostars which may drive jets of high power in comparison to their 
bolometric luminosity. However, taking near-infrared extinction into account as well 
as the fact that ballistic jets tend to be inefficient radiators, the jet powers 
become uncomfortably high.

Another motivating measure concerns the distribution of outflowing mass with radial 
velocity. All of our previous three-dimensional simulations predict shallow 
distributions: a high fraction of gas is accelerated to high speed. In contrast, 
there are numerous outflows for which steep relationships are deduced from CO line 
profiles. We summarise these measurements of the slope in the mass-velocity and 
intensity distributions in \S\,\ref{momentum}. We test here if these problems with 
the efficiency and velocity distribution might both be solved by slowing the rate 
of precession.
\begin{table}
 \caption{Major parameters altered between the simulations discussed (upper set) and the
initial/boundary values for the ambient medium and jet (lower list) which were
not varied.}
 \label{table-pqr}
 $$
 \begin{array}{lccc}
 \hline
 \noalign{\smallskip}
   Run      &\mathsf{P10}  &\mathsf{Q10}  & \mathsf{R10}\\
 \noalign{\smallskip}
 \hline
 \noalign{\smallskip}
 \mathsf{Precession~~ period~~~~} &  50 &  400   &  400 \\
 \mathsf{Precession~~angle} &  10^\circ &  10^\circ   &  10^\circ\\
 \mathsf{Pulsation~~period} &  60 &  60   &  60\\
 \mathsf{Pulse~~amplitude} &  30\% &  30\%   &  0.01\%\\
\noalign{\smallskip}
\noalign{\smallskip}
 \mathsf{Jet~~speed~~(axis)}   &&100~\mathsf{km~s}^{-1} & \\
 \mathsf{Jet~~speed~~(perimeter)}   &&70~\mathsf{km~s}^{-1} & \\
 \mathsf{Jet~~density} && 2.32~10^{-19}~\mathsf{gm~cm}^{-3}  \\
 \mathsf{Jet~~temperature} && 100~\mathsf{K}  \\
 \mathsf{Jet~~radius}  && 1.7~10^{15}~\mathsf{cm} &\\
 \mathsf{Ambient~~density} && 2.32~10^{-20}~\mathsf{gm~cm}^{-3}  \\
 \mathsf{Ambient~~temperature} && 10~\mathsf{K}  \\
 \mathsf{n(H}_2\mathsf{)/n}_\mathsf{H} && 0.5  \\
 \mathsf{Init.~specific~heat~ratio} && 1.4286  \\
\noalign{\smallskip}
 \hline
 \end{array}
 $$
 \end{table}

\begin{figure}
  \begin{center}
  \epsfxsize=8.0cm
    \epsfbox[13 13 161 233]{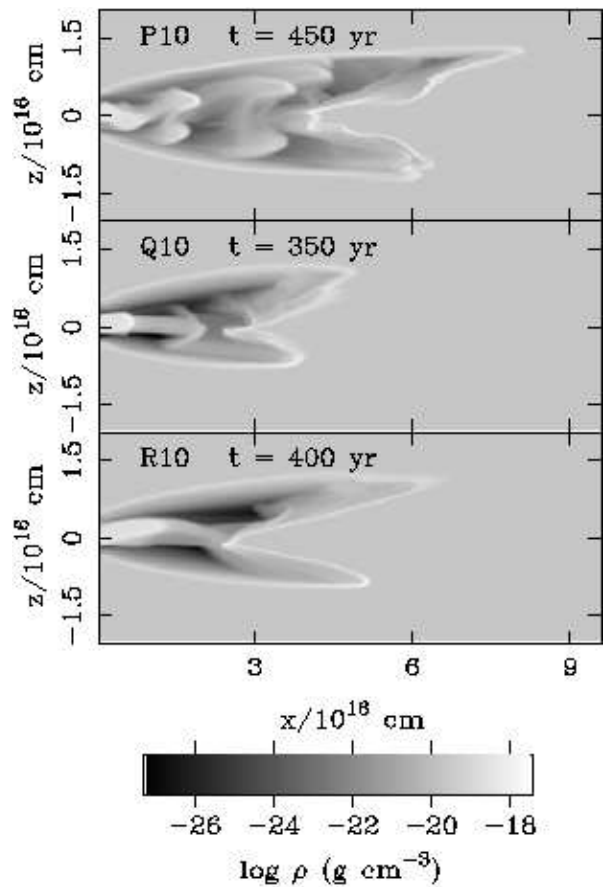}
\caption[]
{Midplane cross sections of density. Each cross section is scaled to the
same maximum and minimum and the scale is displayed below the cross sections.
The complete computational domain in the two axes is displayed. The vertical
axis is the $z$-axis in each panel.  }
\label{allmid}
  \end{center}
\end{figure}

The simulations are executed with a modified version of ZEUS-3D \citep{2003MNRAS.339..133S}
which tracks the molecular fraction, calculates a quite detailed cooling
function, and includes molecular dissociation and reformation. The cooling
and chemistry are solved implicitly. Details and tests
have been presented by \cite{2003MNRAS.339..133S} and \cite{rs1}.

\begin{figure*}
  \begin{center}
  \epsfxsize=16.0cm
  \epsfbox[13 13 144 94]{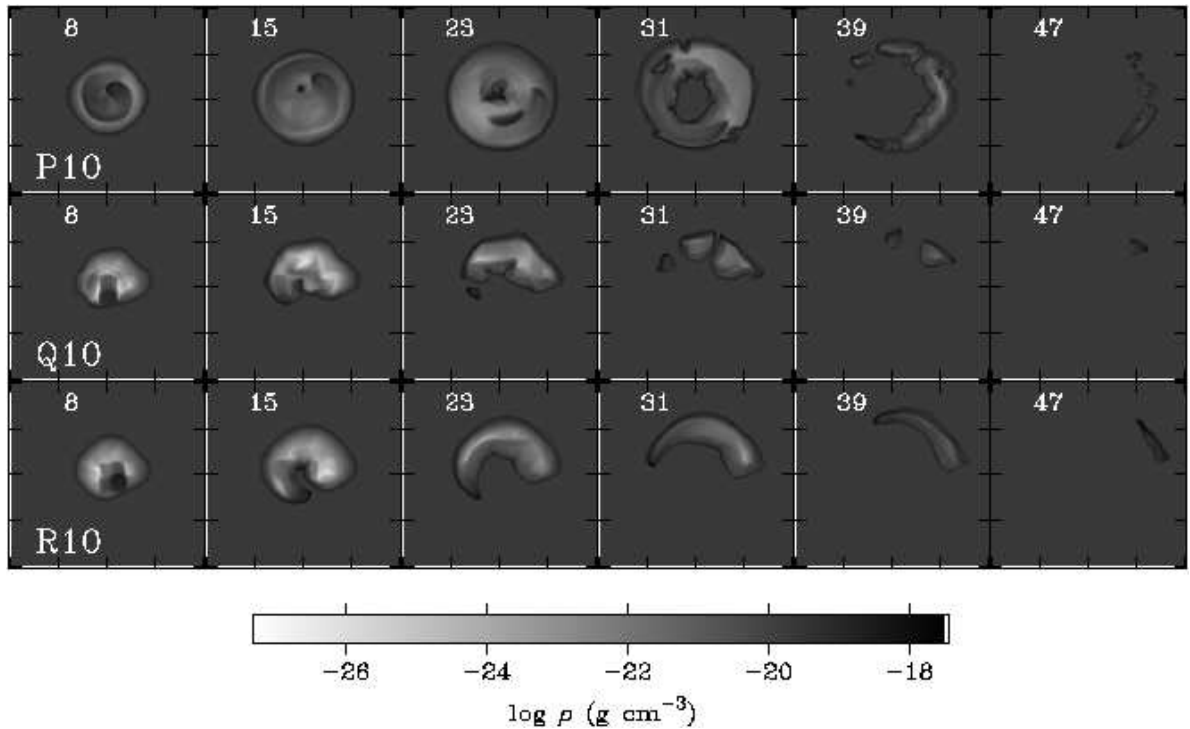}
\caption[]
{Axial cross sections of density for simulations P10 (top row),
Q10 (middle row) and R10 (bottom row). Each row of snapshots is taken from the time
that is displayed in the corresponding panel in Fig.\,\ref{allmid}. The axial position
in $R_{j}$ is given in the upper left of each panel. Each cross section is scaled to the
same maximum and minimum and the scale is displayed below the cross sections. The complete
computational domain in the two axes is displayed. The distance between adjacent tick marks is
1\,$\times$\,10$^{16}$~cm.  The vertical axis is the $y$-axis and the horizontal axis is the
$z$-axis, and so the view is from the jet inlet boundary looking down the $x$-axis
(i.e., toward the +$x$-direction). }
\label{allax}
  \end{center}
\end{figure*}

\section{ZEUS-3D with Molecular Cooling and Chemistry}
\label{method}

\subsection{Hydrocode properties}

We have modified the ZEUS-3D code, which usually updates variables with an
explicit scheme, to include a semi-implicit method to calculate the
time-dependent molecular and atomic hydrogen fractions. We have added a
limited equilibrium C and O chemistry to calculate the CO, OH and H$_2$O abundances.
Equilibrium CO and H$_2$O is a reasonable estimate at high density, consistent
with the low spatial resolution of the cooling layers behind shock waves
\citep{rs1}. The  details of the many components of the cooling function
and the limited chemistry network are discussed in the appendices of
\citet{2003MNRAS.339..133S}.  In summary, we include cooling through rotational and vibrational
transitions of  H$_2$, CO, and H$_2$O, H$_2$ dissociative cooling and reformation
heating, gas-grain cooling/heating, and a time-independent atomic cooling function
that includes non-equilibrium effects. We assume that the gas includes an
additional 10\% by number of helium atoms, so the mean atomic mass is 2.32 $\times$
10$^{-24}$ g. The dust temperature is fixed at 20\,K.

The cooling and chemistry are poorly resolved  as a result of 
the demands placed by the need for three dimensional simulations. The problem is
located within the radiative cooling zones behind shock waves. Since fast
shocks with speed exceeding $\sim$~25~km\,s$^{-1}$ dissociate molecules,
both atomic and molecular cooling zones arise in the present simulations.
In fast shocks, the atomic cooling length is considerably shorter than the molecular cooling length 
and so provides the most stringent scale to resolve. However, this length is of the order of 
just 10$^{13}$~--~10$^{14}$\,cm, given the initial range in density. Therefore,
we make no attempt to resolve the atomic cooling zone and note that peak
temperature obtained in the simulations will be considerably less than the
physical values associated with the shock speed. Fortunately, as utilised in
previous studies \citep{1993ApJ...413..210S,1997A&A...318..595S}, shock conservation 
laws are accurately maintained despite the 
lack of resolution. This is used to argue that the dynamics are correctly
predicted although the atomic emission line fluxes are incorrect (the images of
total atomic cooling presented here should be considered as providing only the general
locations of shocked atomic gas).

The molecular cooling zones are significantly longer: $\sim$ 10$^{20}$/n \,cm. 
While shocks into the lower-density ambient gas are resolved, those in the jet
are generally not. Hence, we restrict our predictions to the properties of the global outflow
rather than individual shocks. This is particularly relevant to the chemistry.
By choosing equilibrium H$_2$O and CO chemistry, the influence of the cooling zone on
the entire flow is minimised. However, a non-equilibrium treatment  of H$_2$ is necessary
and, unless care is taken, this could lead to misleading results. Tests carried out by us
at various levels have shown that the resolution achieved here is adequate
to describe the global properties but that much more detailed fine structure will
appear at higher resolution \citep{1997A&A...318..595S,2002MNRAS.337..477P}.
Other potential  causes of inaccuracy were discussed by \citet{rs2,rs1}.

\subsection{Chosen simulations}

In a previous work, we described a simulation of a molecular jet with a precession angle of
10$^{\circ}$ and a fast precession rate, which we designated P10. In this paper, we discuss P10 and two
additional simulations with a 10$^{\circ}$ precession angle, but with a much slower precession rate
(see Table~\ref{table-pqr}). One simulation is identical to P10 save for the precession rate, and
is designated here as Q10.  The other simulation is identical to Q10 with the
exception of the amplitude of the jet pulsation, which is reduced from
$\pm$30\% to $\pm$0.01\% (essentially a constant initial jet velocity).
We designate this simulation R10.

We modify the jet velocity in P10 and Q10 with a 30\% pulsation and a 60 year period
(see Table~\ref{table-pqr}).  In addition, all simulations have a radial shear that
reduces the jet velocity at the jet edge to 0.7~$v_{j}$.
The precession period is 50 yr in P10, but 400 yr in Q10 and R10. The former is relatively slow
compared to the dynamical time $R_j/v_j$ = 5.4~yr, but fast relative to the
flow evolution time of order 500~yr (see below).  The slower rate is of the
same order as the flow evolution and also our $x$-axis crossing time.

We define the precession rate as fast if it rotates faster than
the outflow expands. Then, the jet contributes to advancing an entire
annulus into the ambient medium, generating a ring structure or annulus of
fixed width (twice the jet radius).
To calculate the expansion time, we first note that the density of the precessing
jet decreases to $\sim \rho_j/(1 + (D \sin \theta)/r_j)$ at a distance D
along the x axis.  This is effectively a one-dimensional expansion since
the width of the annulus remains constant as its
distance from the axis about which the jet is precessing increases.
Momentum balance with the external medium of uniform density $\rho_a$  then yields
\begin{equation}
   \rho_a U^2 \sim  \frac{\rho_j}{1 +(D \sin \theta)/r_j} (v_j - U)^2
\end{equation}
on equating the ram pressures,
where U is the average advance speed of the fast precessing jet into the ambient medium.
If we now consider the expansion to be braked when U is reduced, say, to $0.5~v_j$, then
the expansion  time, $t_E = D/U$, is given by
\begin{equation}
t_E = \frac{\rho_j - \rho_a}{\rho_a} \frac{2 r_j}{v_j \sin \theta}.
\end{equation}

If the precession is slow then we expect a meandering stream to have time
to develop, rather than a ring. For the parameters chosen here,
$t_E = 559$~yr. Therefore, our chosen `slow' precession rate still
displays the effects due to precession
whereas a  `gradual' precession, with a period far exceeding $t_E$ would
approximate to a non-precessing structure.

The computational volume is simulated with a uniform staggered grid divided
into zones each of which spans 2~$\times$~10$^{14}$~cm in each direction.
For all three of the 10$^{\circ}$ precession simulations, we use a
grid of 480 $\times$ 205 $\times$ 205 zones.  Since the jets have large precession
angles, the eventual time-averaged momentum axis for the flow (here, the $x$-axis)
is never the jet flow direction.  The initial jet radius, $R_{j}$, is 1.7 $\times$
10$^{15}$~cm and is therefore resolved by 8.5 zones. The boundary zones satisfy outflow
conditions with the exception of the zones on the inner $x$-boundary that allow
jet inflow.

We assume conditions that we suspect are relevant to Class~0 protostellar jets.
In all cases, the jet is initialised with a hydrogenic nucleon density
of 10$^5$ cm$^{-3}$.  We set the jet-to-ambient density ratio
to 10, the initial jet temperature to 100\,K and the ratio of
jet-to-ambient thermal pressure to 100.  The nominal, mean axial jet speed is
$v_j = 100$~km~s$^{-1}$.  This yields an initial Mach number with respect to
(the uniform) ambient medium of 140.

\section{Physical Properties}
\label{properties}

\subsection{Internal structure}

Midplane slices of density are displayed in Fig.\,\ref{allmid}
at a simulation stage where side projections of the outflows almost spanned
the entire computational grid. In fact, at the times shown in Fig.\,\ref{allmid},
both Q10 and R10 have jet material as far from the inlet as $x \sim
8.5 \times 10^{16}$ cm (50 R$_j$).  Most striking is the existence of a jaw-shaped structure
in the slow precessing cases Q10 and R10 as well as the fast precessing case.
Also, slow precession generally yields lower density cavities (darker in Fig.\,\ref{allmid})
since swept up material is not quickly replaced by the jet.  The midplane density
slices reveal the shocks from the pulsing in Q10, and relative smoothness
of the non-pulsed flow in R10.

We display the cross-sectional structure normal to the jet axis for the
same three simulations in Fig.\,\ref{allax}.
For P10, this shows a complete annulus mid-way down the jet which
fragments near the leading edge.  For Q10 and P10, the slow precessing
results in a helical flow pattern from $+z$ (the right side of each panel)
to $-z$ through $+y$ (the top of each panel) and back again through $-y$.  This is
consistent with the time-variation of the precession at the origin. A
comparison between the Q10 and R10 jets reveals that the instability seen in
the ``islands" of jet material at large $x$ are primarily caused by the
pulsing.  Specifically, the islands seen in Q10 at $x/R_j$ = 31 and
39 are unified into an arc-like bow structure in R10.
However, as presented below, the integrated H$_2$ emission portends future
fragmentation in the R10 flow.

\begin{figure}
  \begin{center}
 \epsfxsize=8.0cm
    \epsfbox[85 400 430 775]{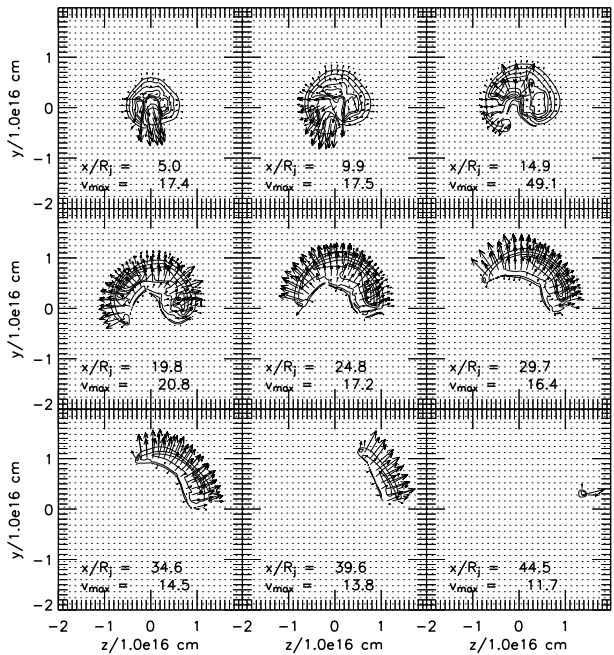}
\caption[]
{Transverse velocity vectors and mass density contours in simulation R10 at t = 350\,yr.
The value labelled $x$ is the axial position and v$_{max}$ is the maximum transverse velocity
in the panel (in km~s$^{-1}$).  The velocity vectors are displayed for every sixth
zone and the maximum vector length in any panel is normalised to the maximum
velocity in that panel.  Density contours are at log~$\rho$ = -24.0, -23.0, -22.0, -21.0, -20.0, -19.0,
and -18.0.}
\label{velvec}
  \end{center}
\end{figure}

Quite high transverse velocities are encountered in the slow precessing cases.
A plot of velocity vectors and density contours (Fig.\,\ref{velvec})
for R10 shows that outward facing velocity vectors are associated
with the helical bow shock far from the inlet.  Closer to the inlet
(in the first couple of panels), the precessing jet is evident.  In addition,
there is a large transverse velocity ($\sim$ 50~km~s$^{-1}$) in
the third panel, where the bow shock has nearly spun around on itself.

\subsection{Speed and size}   

We have previously analysed the fast precessing jet simulations to determine their
capability of producing giant outflows of size $\sim$~1~pc \citep{rs3}. We employed a 
model to fit the mean speed of advance and to determine the effective drag coefficient
corresponding to the aerodynamic properties and atmospheric resistance.
This term, C$_d$, can also be parameterised as a change in area across the bow shock
in the usual ram pressure approximation for the Mach disk advance speed, as used in
\S\,\ref{method}.  We deduced low values of C$_d$ in the range 0.18--0.27 for a range 
of density ratios and mass fluxes whereas we expect a value of unity if the ram pressure 
of the jet is modified by its expansion only.

For the fast precessing cases, the drag coefficient increases with precessing angle, from 
0.20 for 5$^\circ$ half-precession angle to 0.45 for the 20$^\circ$ case.  Within this model, 
we found that the fast precessing jets could reach a distance of 1~pc between 35,000 and 
85,000 yr, showing evidence for deceleration despite the small drag coefficients. We concluded 
that outflows reach large sizes in times comparable to the Class~0 protostellar stage.

\begin{figure}
  \begin{center}
  \epsfxsize=7.8cm
\epsfbox[0 150 430 500]{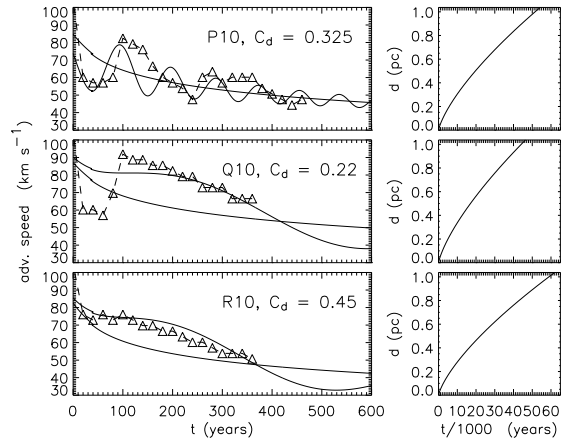}
\caption[]
{Advance speed and estimated time to 1~pc for precessing
molecular jets.  The panels on the left show the time-averaged
(over 20 years) advance speed of the bow shock, which
are indicated by the triangles connected with the dashed line.  The advance speeds have
been fitted with a model that includes the momentum drag from the interaction of the
jet with the ambient medium \citep{rs3}.  The inferred drag coefficient,
C$_d$, is listed in the
upper right of these panels.  The models are indicated
by the solid lines, the most accurate includes the slowly-damped sinusoidal variation
caused by the precession while the monotonically decreasing fit assumes
that the jet momentum is evenly spread in the advancing disc/annulus. The panels on the right
show the time required to reach 1~pc, according to the extrapolation of the advance speed.
}
\label{advancespeed}
  \end{center}
\end{figure}

Here, the evolutions of the linear extents are shown in Fig.\,\ref{advancespeed}. Also shown are 
the best fit models and integrated times to 1~pc. We find that the drag coefficient varies 
according to the physical structure. The advancing bow shocks of run Q10 result in a low 
drag coefficient of 0.22. In contrast, the continuous ribbon produced by run R10 leads to the highest 
drag (0.45). This is consistent with expectations from  fluid dynamics since the 3D 
aerodynamic bow shapes permit fluid to flow easily past the obstacles, whereas the ribbon 
presents a more obtrusive object. The time required  to reach 1~pc according to the 
extrapolated model remains within the 40,000--60,000~yr range.

\begin{figure}
  \begin{center}
 \epsfxsize=8.0cm
    \epsfbox[13 13 157 202]{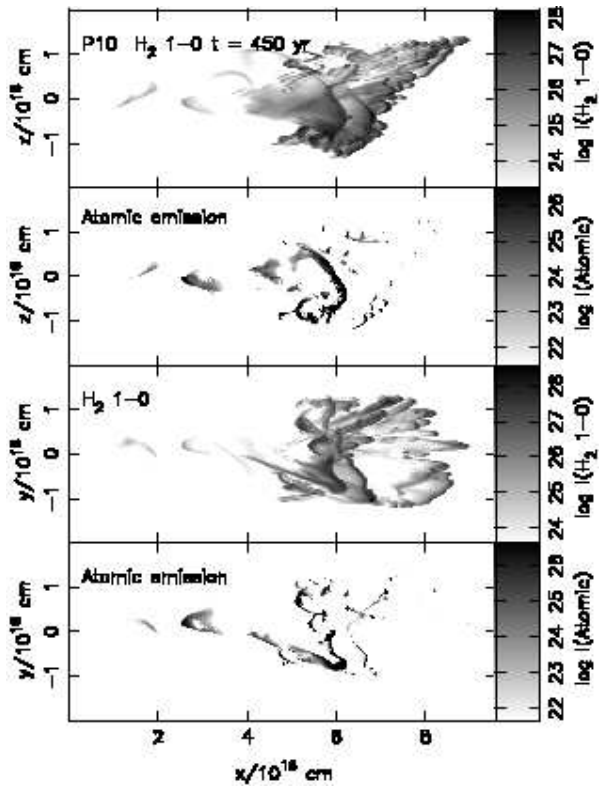}
\caption[]
{Two orthogonal views of simulation P10 at $t$ = 450~yr in one molecular emission line
and one general ``atomic'' emission line.  In the top two panels, the
integration is along the $y$-axis, and in the bottom two, integration is along
the $z$-axis.  In order to produce a more complete picture of atomic emission, the image
has been rescaled so that the maximum luminosity in the appropriate panels is 3
orders of magnitude lower than the actual maximum luminosity (which is very localised).
The luminosities are in erg~s$^{-1}$ from bins with roughly the same size as the 3D zones used in the
simulations (i.e. 2\,$\times$\,10$^{14}$ cm).}
\label{imageyz-p10}
  \end{center}
\end{figure}

\section{Simulated Molecular Line Emission}

\subsection{Integrated line emission}

\begin{figure}
  \begin{center}
 \epsfxsize=8.0cm
    \epsfbox[13 13 168 252]{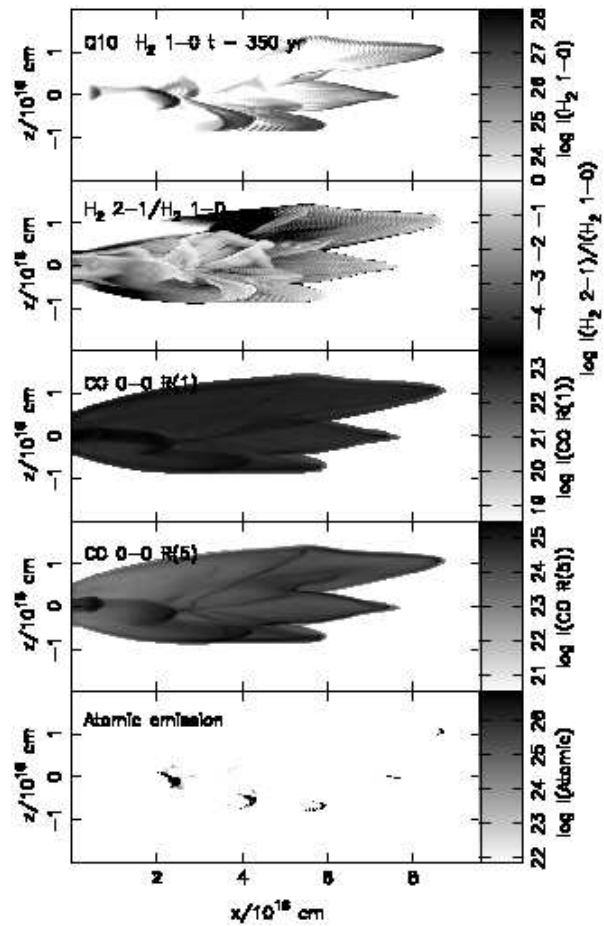}
\caption[]
{Integrated emission along the $y$-axis in simulation Q10 at $t$ = 350 yr for four molecular 
emission lines and one general atomic emission line. Further details are as 
in Fig.\,\ref{imageyz-p10}. }
\label{imagez-q10}
  \end{center}
\end{figure}

\begin{figure}
  \begin{center}
 \epsfxsize=8.0cm
   \epsfbox[13 13 140 213]{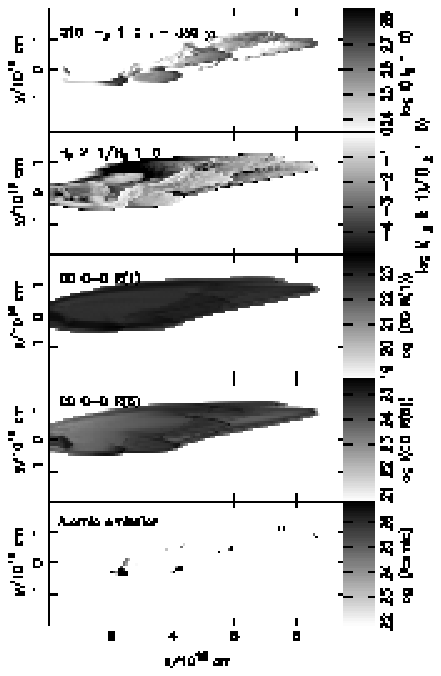}
\caption[]
{Integrated emission  along the $z$-axis from simulation Q10 at $t$ = 350 yr in four 
molecular emission lines and one general atomic emission line. Further details are as 
in Fig.\,\ref{imageyz-p10}.}
\label{imagey-q10}
  \end{center}
\end{figure}

\begin{figure}
  \begin{center}
 \epsfxsize=8.0cm
    \epsfbox[13 13 164 246]{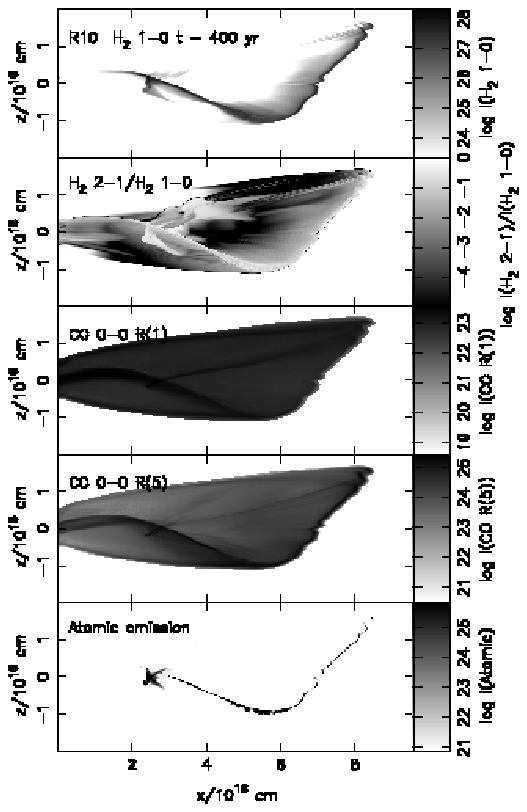}
\caption[]
{Integrated emission along $y$-axis from simulation R10 at $t$ = 400 yr in four molecular 
emission lines and one atomic emission line.  Further details are as in
Fig.\,\ref{imageyz-p10}.}
\label{imagez-r10}
  \end{center}
\end{figure}

\begin{figure}
  \begin{center}
 \epsfxsize=8.0cm
    \epsfbox[13 13 139 205]{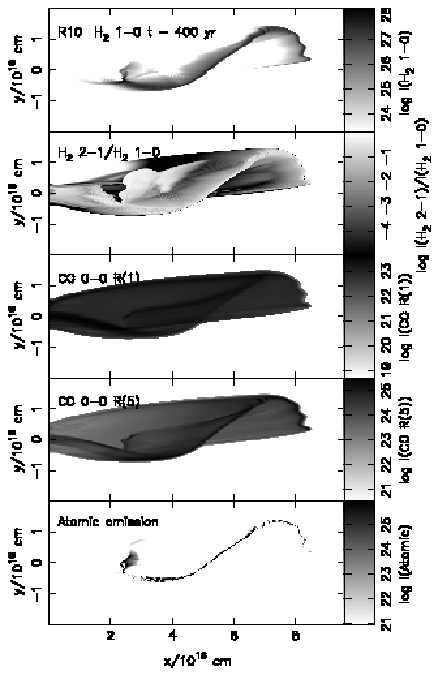}
\caption[]
{Integrated emission  along $z$-axis from simulation R10 at $t$ = 400 yr in four molecular 
emission lines and one atomic emission line. Further details are as in Fig.\,\ref{imageyz-p10}.}
\label{imagey-r10}
  \end{center}
\end{figure}

\begin{figure}
  \begin{center}
  \epsfxsize=8.0cm
\epsfbox[13 13 108 140]{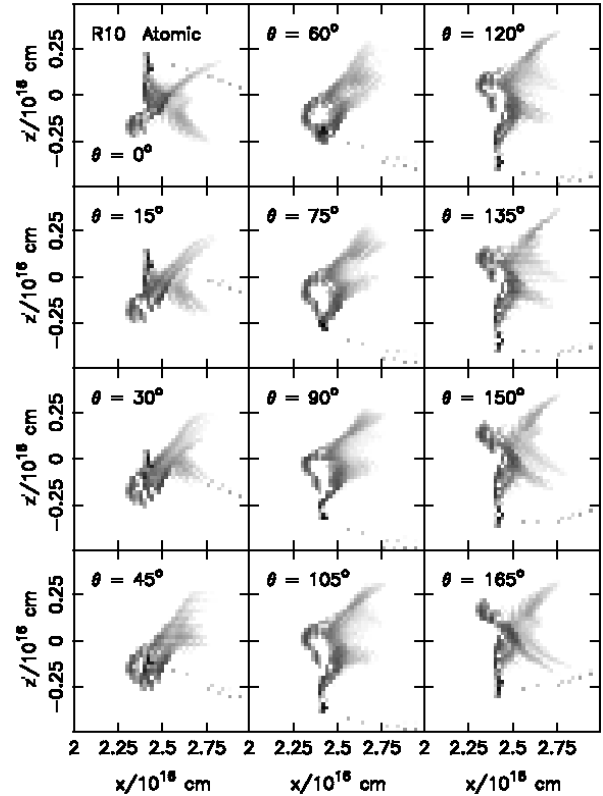}
  \caption[]
{Close-up rotation of views  about the axis in simulation R10.
By rotating about the jet axis of the ``x" feature in Fig.\,\ref{imagez-r10},
we see a more varied appearance.
 }
\label{xanim}
  \end{center}
\end{figure}

\begin{figure}
  \begin{center}
  \epsfxsize=8.0cm
\epsfbox[11 13 154 127]{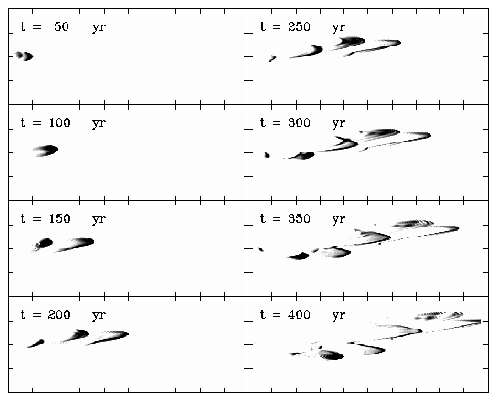}
  \caption[]
{Evolution of the 1--0~S(1) molecular hydrogen emission for simulation Q10.  Integrations
of H$_2$ 1--0 emission along the $z$-axis (the $y$-axis is the vertical
dimension in all the panels). }
\label{anim}
  \end{center}
\end{figure}

We  estimate the associated radiated energy rate for a few molecular
and atomic emission lines using the zonal values for mass density, temperature and molecular
fraction.  We integrate the emission for selected molecular lines
at specific times with the assumption that the outflow and
foreground cloud are optically thin; examples of these integrations
are shown in Figs.\,\ref{imageyz-p10}--\ref{anim}.
The H$_2$ line emission is based on a non-LTE
approximation to the vibrational populations \citep{1983ApJ...264..485D} and the
CO rotational emission is calculated according to \cite{1982ApJ...259..647M}.
The only plot presented for run P10 (Fig.\,\ref{imageyz-p10}), which was discussed previously,
includes new integrations in ``atomic" emission as well as in molecular
hydrogen emission for comparison.

The atomic emission is derived from the cooling function used in the code (based on
\citealt{1993ApJS...88..253S}), unmodified to select a specific emission line.
As with all of these integrations, they provide a guide to the location and
extent of emission and relative brightness of regions of the flow rather than accurate
predictions.  In run P10, the brightest atomic emission is spatially correlated with 
relatively bright regions of molecular hydrogen emission, a result that is not 
surprising since both are related to the cooling of post-shock material. However, 
the brightest atomic emission occurs in localised small knots and filaments whereas 
the H$_2$ emission is relatively more diffuse.

The images of atomic emission prominently display inner cones associated with a
stagnation point (the knot at 3 $\times$ 10$^{16}$ cm and some of the emission
forward of that).  Also, there is no extended atomic emission from the leading
edge of the jet, where there is significant H$_2$ emission. This suggests that the 
atomic emission is produced in strong shocks directly  associated with the jet impact 
rather than the predominantly weak oblique shocks associated with the advancing
aerodynamic bow-shaped features.

The simulated images of molecular hydrogen and atomic emission reveal the shock structure. 
In simulation Q10 (Figs.\,\ref{imagez-q10} and \ref{imagey-q10}), the pulsing jet at the 
inlet leads to many internal shocks, while in R10 (Figs.\,\ref{imagez-r10} and \ref{imagey-r10})
the constant speed yields a shock in the form of a single sinuous ribbon.

This ribbon contains an apparent ``x" at $x \sim 2.5 \times 10^{16}$ cm in the $y$-integrated 
images (Fig.\,\ref{imagez-r10}). It is most prominent in
the atomic emission from simulation R10 but  is also
bright in all emission lines shown.  While more difficult
to perceive, a similar feature is also distinguishable in the plots
for run Q10.  In the images generated by $z$-integration  (Fig.\,\ref{imagey-r10}),
this ``x" is a more complicated structure, suggestive of a
disk of emitting material seen edge-on.  Such a disk
is confirmed by viewing the jet from various angles about the mean momentum ($x$) axis,
which we display in Fig.\,\ref{xanim}.  Moreover, this disk can be seen in
in the second panel of the R10 density cross sections in Fig.\,\ref{allax}
and contours of the third panel in Fig.\,\ref{velvec}.

From these images, we hypothesise that the ``x" is related to a
stagnation disk that at the time of this image is being
squeezed from both sides, since the precession rate is
just sufficient for more than a half-turn to exist at
this $x$-position.  In addition, the third panel in the
velocity vector plot has a very large maximum transverse velocity
($\sim$ 50 km s $^{-1}$ compared with 17.5 km s$^{-1}$ transverse
velocity from the nominal jet speed and the input angle).
It seems plausible that this large velocity is caused by
the release of accumulated material from this stagnation
disk.  To investigate further,
we have also executed isothermal simulations of a fast dense jet which show
evidence for large tangential velocities near the jet-crossing point.

The ``x" in atomic emission appears similar to that in the HH\,222
(seen in a plate of the HH34 superjet complex, \citealt{1994ApJ...428L..65B}),
although the the process responsible is likely different in the real source,
which is many times the size of what has been simulated here.

We present the temporal variation of molecular hydrogen emission from
run Q10 in Fig.\,\ref{anim}.  As opposed to the fast precessing cases, where
at early times we saw an elliptical ring form, here we see a
strong asymmetry about the mean momentum axis.  However,
at early times this axis may not be obvious.  Additionally,
the bow shocks evolve in a complicated and fascinating fashion,
with the strongest emission moving forward through the bow
as the entire emitting region moves forward.  For times just after 
t = 300 yr, there is a rapid change in the direction of the jet,
which could be (mis-)interpreted as a dramatic event
at the jet origin.  Of course, in this case we know that only
a smooth variation is responsible.  The sequence of bow shocks at late
times bears a striking similarity with L1634 \citep{2004A&A...419..975O},
which is an order of magnitude larger than our simulations.

The vibrational excitation is measured by the H$_2$ 2--1/1--0 line ratio.
The excitation distributions reveal the most intense shocks, and its brightest regions
generally coincide with bright regions of the 1--0 line itself.  Noteworthy
is that the region associated with the ``x" feature in R10
has quite a high H$_2$ line ratio.  

The moving CO emission displays an outline of the dense jet plus swept-up ambient material.
The CO emission lines, particularly the 0--0 R(1), emphasise the sinusoidal projection
of the helical flow pattern. The ribbon is traced back to the source. Note that the
individual advancing bow shocks in run Q10 sweep out separate protruding CO cavities
(Fig.~\ref{imagez-q10}). Finally, the CO R(5) images (i.e. in the rotational emission line
with a frequency of 691\,GHz, from 116\,K above the ground state)  
again appear somewhere between the H$_2$ emission and the CO R(1), as expected. Both the
cavity structure and internal bullets are evident. 

\subsection{Total line emission and efficiency}
\label{totalline}

The evolution of the total molecular line emission from molecular outflows
is of relevance to the global modelling of star formation (e.g., the Unification Scheme,
\citealt{2000IrAJ...27...25S}; \citealt{2000PhDT........12S}; and \citealt{2000AJ....120.1974Y}).
In contrast with previous pulsating and some of the fast precessing simulations,
the total molecular intensities for molecular hydrogen and CO monotonically increase
over time for the new simulations presented here.  They cover the same range
as run P10 (see Fig.\,10 in \citealt{rs3}).  The non-pulsing case R10
increases in the most smooth fashion of the three simulations discussed here.
The periods of flat or decreasing luminosity that exist in Q10 coincide with
the appearance of a new knot near the inlet.  Eventually this knot
brightens sufficiently that the total flux increases.

We have also calculated the ratio of the total emission compared
to a time-averaged kinetic energy flux (mechanical power) from the
jet source for a sample of our molecular jet simulations (see Fig.\,11 from
\citealt{rs3}).  Typically, the H$_2$ 1--0 emission line had an
efficiency of around 0.1--1.0\% of the mean mechanical energy,
with an increase in this range over the lifetime of the simulation.
For CO the range varied from 10$^{-7}$ at early times to roughly 10$^{-5}$
at late times ($t \sim 500$ yr), with a more
monotonic variation with time between the two extrema than in the
H$_2$ case.  We have also computed the evolution of these ratios for
the two new simulations and find values close to
that of Simulation P10.  In general, both the variation over time
in the mechanical power and in the H$_2$ to mechanical power
ratio (by 0.2 dex, usually) is smaller in the slow precessing cases
than in P10.  Also, the ratio is typically smaller (0.1 dex) in the pulsed
case (Q10) than non-pulsed case (R10).  For the CO ratio, Q10 is smaller or equal
to the ratio in P10, and the R10 CO ratio is typically larger or equal to that of
P10, but over time both asymptotically approach the evolution of P10.

We have also calculated the the instantaneous radiative loss from each part of our cooling function 
at 50 yr intervals for the new simulations.  These
losses are then compared with the time-averaged mechanical luminosity (power) input at the inlet
up to the point in time that the losses are computed.  We find that at t = 350 yr, the total
losses are roughly 30\% of the mean power, with two-thirds of that being
from a combination of H$_2$ ro-vibrational and atomic cooling.  For the
pulsed case (Q10), the atomic emission alone is about 5\% of the mean power
at this time, but for the non-pulsed case (R10), it's fraction is only 0.5\%.
This is explained by the many internal bow shocks, which are sites of heating
and dissociation for the initially molecular gas.  
The evolution of the H$_2$ 1-0 S(1) emission efficiency from 0.1\% to 1\% in
these simulations is associated with a nearly constant 4\% ratio of emission from this line
to the total H$_2$ ro-vibrational cooling rate.

Other major contributors to the cooling function include (values in parentheses are the fractions of
each cooling component of the mean power for the jet near the end of the
simulation): H$_2$ dissociation (2--3\%), H$_2$O rotational
transitions due to collisions with H and H$_2$ ($\sim$ 5\%), and CO rotational
transitions due to collisions with H and H$_2$ (2\%).
The individual component ratios of luminosity to mean jet power nearly always increase 
monotonically with time, as does the total, which increases by about 5\% of the mean power in 
every 50~yr interval.



\section{Velocity Distribution of Mass and Molecular Line Emission}
\subsection{Background}
\label{momentum}

There have been many recent measurements of the slope in the mass-velocity
distributions, interpreted from CO line profile data of
protostellar outflows. Once the slow turbulent motions of nearby molecular
clouds are meticulously subtracted (e.g., \citealt{2000AJ....120.1974Y}), there is
frequently a quite shallow distribution (with
$\gamma$ in $dm/dv \propto v^{-\gamma}$ occasionally as low
as 1) at low radial velocities. A steeper slope ($\ga$ 6) is found at higher velocities, where the turnover
occurs at $v_{break} \sim$ 10 km~s$^{-1}$.  Note, however, that the majority of powerful outflows
may possess very steep profiles even at low speeds \citep{1998AJ....115.1118D,2001A&A...378..495R}.
Many of these outflows possess low collimation, suggesting that the
wide outflow sweeps up a high mass to low speeds.

Additionally, a similar analysis has been performed for H$_2$ line profiles of some
protostellar outflows \citep{2002ApJ...572..227S}. The
equivalent  $\gamma$ is defined as $\gamma_m = dL(1-0\,S(1))/dv \propto v^{-\gamma_m}$.
The authors measure a range of $\gamma_m$ between 1.8 and 2.6
above  a break velocity which lies between 2 -- 17 km s$^{-1}$ (the flux is flat or
slightly rising below this radial velocity). We wish to test here if jets undergoing wide-angle
long-period precession could decelerate sufficiently to match the observations of large
$\gamma_m$ at high speeds.

Previous simulations of ballistic jets with small precession angles
yielded quite shallow values for the CO $\gamma$ of 1.2\,--\,1.6
\citep{1997A&A...323..223S} or even shallower \citep[0.9\,--\,1.3,][]{rs1}.
Non-ballistic jets, with equal or lower density than the ambient medium, also
generated outflows with low  $\gamma$ values \citep{rs1}.
The fast precessing jets in \cite{rs3} yielded a somewhat larger range
of slopes for the small velocity region, (0.4\,--\,1.6).  The higher $\gamma$
values are associated with combinations of the following conditions:
jets viewed near the plane of the sky, jets with a large precession
angle, or jets which are initially atomic.

The simulations in \cite{rs1} possess flux-velocity distributions of H$_2$
that do not closely resemble the observations of \cite{2002ApJ...572..227S},
with a strong increase in luminosity at small velocities. This is not
disturbing since the 1-0~S(1) emission properties are very sensitive to the physics
within the radiative shocks. In hydrodynamic shocks, the molecules radiate
efficiently only after being strongly accelerated and compressed.
This suggests that the shock viscosity is instead  provided by ambipolar diffusion
in the molecular sections.

Axisymmetric simulations similar to those in \cite{1999A&A...345..977D} were
analysed by \cite{2003dc}. They obtained a low-velocity CO $\gamma$ of
1.5, which is in the range of commonly observed values.  This may have been the result of
smaller jet densities and longer time coverage than we have used.  Indeed,
\citet{1997A&A...323..223S} also found that the $\gamma$ value increases with
time in 3D simulations.  \cite{2003dc} also attempt to
reproduce the H$_2$ results of \cite{2002ApJ...572..227S},
with some success but  use an approximation (LTE)
that is more appropriate for densities larger than in most of
their flow.

\subsection{Mass- and intensity-velocity relations}

\begin{figure}
  \begin{center}
  \epsfxsize=8.50cm
 \epsfbox[75 30 580 630]{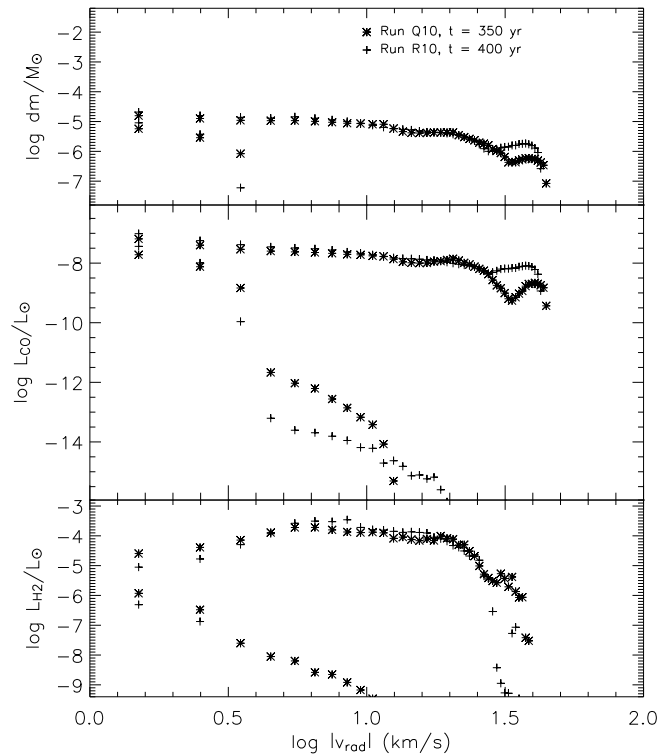}
\caption[]
{ Distribution of velocities into bins of mass and two molecular line luminosities for a viewing
angle of 15$^{\circ}$.  The distributions of mass (top), CO luminosity (middle) and H$_2$ 1--0 emission
(bottom panel) are shown. Each velocity bin is 1~km~s$^{-1}$ wide.
The designation for the data presented in each panel is shown in the top panel.
Each run may be represented twice within each panel, with data from both positive and
negative radial velocities included.  Naturally, the lower valued data is for the
positive radial velocities (which could contribute to a  ``red" lobe, while the other is from a ``blue"
lobe). }
\label{gamma}
  \end{center}
\end{figure}

We have performed a similar analysis for the velocity distributions of mass,
CO R(1) luminosity and  H$_2$ 1--0 luminosity for each of the simulations
presented here. Power law slopes $\gamma$, $\gamma_{CO}$ and  $\gamma_{H2}$ are determined.
We list the results for a sample of viewing angles, defined as the angle of the $x$-axis out
of the plane of the sky toward the observer, in Table~\ref{gammas}.  We also display the
data for one viewing angle (15$^{\circ}$) in Fig.~\ref{gamma}. Note that both red and
blue shifted line wings are displayed in this figure.  A summary of our results
follows.

\begin{table*}
     \caption[]{Mass Spectra Power-Law Dependences}
     \label{gammas}
     \begin{tabular}{lccclllcclll}
          \hline
          \noalign{\smallskip}
 {\rm Sim.} &  t(yr) & type  & $\theta$ & range of log $|v|$ & $\gamma$ & N$^{\rm a}$ & type & $\theta$ & range
of log $|v|$ & $\gamma$ &  N$^{\rm a}$\\ \noalign{\smallskip}
\hline
\noalign{\smallskip}
P10        & 450 & mass & 15 & 0.5--1.0/1.0--1.4 & 1.02/1.87 & 7/15 & CO & 15 & 0.5--1.0/1.2--1.4 & 1.20/1.96 &7/9 \\
            &         &         & 30 & 0.5--1.0/1.1--1.7 & 0.52/1.68 & 7/37 &       & 30 & 0.5--1.0 & 1.22 & 7 \\
            &         &         & 45 & 0.5--1.0/1.4--1.8 & 0.63/1.65 & 7/38 &       & 45 & 0.5--1.0 & 1.22 & 7 \\
           &         &         & 60 & 0.5--1.0/1.4--1.9 & 0.57/1.45 & 7/54 &       & 60 & 0.5--1.0 & 1.21 & 7 \\
           &         &         & 90 & 0.5--1.0/1.6--1.9 & 0.46/1.37 & 7/39 &       & 90 & 0.5--1.0 & 1.21 & 7 \\
\noalign{\smallskip}
Q10 & 350 & mass & 15 & 0.5--1.0 & 0.26 & 7 & CO & 15 & 0.5--1.0 & 0.49 & 7 \\
            &         &         & 30 & 0.5--1.0 & 0.56 & 7 &      & 30 & 0.5--1.0 & 0.76 & 7 \\
            &         &         & 45 & 0.5--1.0 & 0.63 & 7 &      & 45 & 0.5--1.0 & 0.80 & 7 \\
            &         &         & 60 & 0.5--1.0 & 0.59 & 7 &      & 60 & 0.5--1.0 & 0.74 & 7 \\
            &         &         & 90 & 0.5--1.0 & 0.53 & 7 &      & 90 & 0.5--1.0 & 0.65 & 7 \\
\noalign{\smallskip}
R10 & 400 & mass & 15 & 0.2--0.7/0.8--1.3 & 0.35/1.12 & 3/14 & CO & 15 & 0.2--0.7/0.8--1.3 & 0.83/1.35 & 3/14 \\
            &         &        & 30 & 0.5--1.0 & 0.63 & 7 &       & 30 & 0.5--1.0 & 0.71 & 7 \\
            &         &        & 45 & 0.5--1.0 & 0.72 & 7 &       & 45 & 0.5--1.0 & 0.79 & 7 \\
           &         &         & 60 & 0.5--1.0 & 0.65 & 7 &       & 60 & 0.5--1.0 & 0.76 & 7 \\
           &         &         & 90 & 0.5--1.0 & 0.56 & 7 &       & 90 & 0.5--1.0 & 0.70 & 7 \\
\noalign{\smallskip}      \hline
\end{tabular}
\begin{list}{}{}
\item[$^{\rm a}$] N is the number of points in the velocity distribution, which was computed in 1 km s$^{-1}$ bins,
used to estimate the slope.
\end{list}
\end{table*}

\begin{itemize}
\item The precession period has a small influence on the mass distribution as
measured by the power-law index $\gamma$.  For viewing angles $> 15^{\circ}$, there is a slightly steeper 
slope in the mass relation in the slow-period precessing cases, which have slopes that are fairly close to
each other (the non-pulsed case marginally increases the slope).
\item For the CO distribution, there is a stronger and opposite dependence of
$\gamma_{CO}$ on the precession rate, with most viewing angles of the slow-precession cases
yielding a slope 0.5 smaller than the rather shallow 1.2 for P10.
\item In simulations of fast precessing jets, we found that the slopes for
the CO distribution exceed those of the mass distribution, occasionally quite significantly.
This is similar to a result from the low density jet simulations of \cite{2003dc}. This
trend extends to the slowly precessing molecular jet simulations presented here,
although the differences between the two slopes are small in Q10 and R10 than
in P10.
\item At least for large viewing angles ($>$ 45$^{\circ}$), we confirm previous results that there is
an inverse relationship between $\gamma$ and the viewing angle (\citealt{1997A&A...323..223S},
\citealt{2001ApJ...557..429L}, \citealt{2003dc}, \citealt{rs1} and \citealt{rs3}).
\item Additionally, we have examined the evolution of the velocity distributions for each of the
molecular jet simulations at two different programme times.  In contrast to the findings of
\cite{1997A&A...323..223S}, we find that the slope of these
distributions {\it shallow} slightly over the small length of time simulated here.
\item The velocity distributions of H$_2$ emission for the slow precessing cases contrast with the
observationally determined ones, as with our previous simulations.  Instead of the observed flat
distribution at low velocities, we find a dramatic rise.
\end{itemize}

\subsection{Discussion of position-velocity images and channel maps}

In previous papers, we have presented position-velocity (P-V) diagrams in molecular emission
lines from various simulations.  From simulations of slow precessing jets,
we find no additional features and so omit any new figures here.

However, there are some differences.  The main one is that in the middle of the projected jet
(x$^{\prime}$) axis, the H$_2$ diagram appears much less ``turbulent", even in the Q10 pulsed case,
than in P10 and thus the flow is much better focussed for the slower precession rate.
In the R10 case, the middle portion
is contained completely within 0--10 km s$^{-1}$, although this is not true for the CO emission, which
extends out to 20 km s$^{-1}$.

In R10, rather than the sequence of Hubble Law like P-V regions (that are
a consequence of the pulsing) we find an arc of bright material that starts at high velocity
(35 km s$^{-1}$) near the origin and swings down to  10~km~s$^{-1}$ near the middle of the plot.
About half-way along (from about one-quarter the jet length in x$^{\prime}$ out to the maximum
extent of the jet), the entire range of velocities out to the arc are filled in.
In addition, the maximum velocity increases from the middle portion outward, to 25 km s$^{-1}$.

Analysis of previous velocity channel maps, primarily in CO, revealed a morphological
progression as the viewed velocity is changed.  High velocity maps show the isolated
knotty structures associated with internal shocks along the jet and low velocity maps
display the full extent of the shocked ambient medium.  In the fast precessing cases,
the knots in the high velocity maps are displaced from the mean momentum axis, frequently
in an alternating pattern.  In the slow precessing, pulsing jet case (Q10), we
see a sequence of alternating knots, but the spatial frequency of the oscillation
has decreased.  In both Q10 and R10, the velocity channel maps indicate that
the bright arc within the CO emission near the inlet is at high velocities,
while much of the integrated CO emission (Figs.\,\ref{imagez-q10} and \ref{imagez-r10})
is at low velocities.

\section{Summary of Results}



We have found that the dominant physical structure for all precessing jets
is an inward facing cone. In other words, the jet clears out
a similar volume of ambient gas. The jaw-shaped structure contrasts with the
tube-like structure expected if the jet were to be bent or deflected.

A fast precessing jet generates an impact ring which fairly rapidly disrupts into many
mini-bow shocks.  Here, slowly precessing jets lead to  helical flows
that remain noticeably more stable for the duration of the simulation.

A slow precessing and pulsating jet cuts through the ambient medium with the
highest speed. We measure this here through an effective drag coefficient.
The non-pulsating jet results in a drag coefficient about double that of
the pulsating case. Therefore, bow shocks represent the most efficient means
of advancing an outflow yet the least efficient at transferring momentum.

The stagnation region at the apex of the inward facing cone displays
quite complex time-dependent behaviour in the slow precessing cases.
The feature does move slowly
forward and accumulates some material, which after some time
can be sloughed off to the side open to the ambient medium.  We believe
that this is the cause for the high transverse velocities measured in these
simulations (particularly the non-pulsed R10 case).  The feature, which
at times can resemble a disk in 3D, is seen in atomic and high energy
molecular emission as an ``x" feature from certain viewing angles.

The images of molecular emission show quite dramatic evolution.  In the pulsed 
case, the H$_2$ emission shows a train of asymmetric wedge-shaped bow shocks. 
The sequence of bow shocks 
can appear over time to quickly change direction suggesting a violent event at 
the driving source, yet we are simulating a smoothly varying
jet flow. The distribution and structure of the bow shocks appear very similar to the
observed HH\,240A--D system in the L\,1634 outflow \citep{2004A&A...419..975O}.
Note, however, that the HH\,240 bows have advanced a distance of $\sim 0.4$\,pc 
from the driving source. Many other groups of bows have been uncovered (e.g.
HH\,195 and HH\,197 5 \citep{2000A&A...354..236E}). 

On the other hand, shock emission in the form of ribbons is rarely encountered. 
Examples are HH\,222 \citep{2002AJ....123.2627B}, HH\,103--105 
\citep{2000A&A...354..236E} and IRAS\,23151+5912 \citep{2005weigelt}. In fact, HH\,222 also possesses an  ``x" feature,
resembling that predicted in our simulations.

The CO cavity patterns found here are largely featureless. In the
pulsed case, however, distinct shells may outline the wakes of separated bow shocks
which advance directly into the ambient cloud. To observe this property
requires resolution and sensitivity which may soon be generally attainable.
At present, only one well-known precessing outflow, RNO\,43, is known to possesses this
striking property \citep{1996MNRAS.279..866B}.

Many bipolar outflows contain a high fraction of cool gas moving at radial velocities
near to that of the embedded cloud. This is derived from CO intensity-velocity 
relationships which are approximated by a high negative power-law index.
In contrast, low power-law indices are associated with these simulations.
Hence, the solution to the discrepancy between our results and observations
lies elsewhere. It might be reconciled by running the simulations
for longer periods of time, which should be easier to perform as faster and
larger computing facilities become available.
Despite the lack of resolution some interesting outstanding issues, we have shown
that the precession rate can dramatically influence the appearance of precessing 
molecular protostellar jets.

The simulations ignore the effects of the magnetic field. Simulations under
ideal magnetohydrodynamic conditions may show significant differences. For example,
the magnetic field may stabilise the compressed layer produced by the advancing 
shock wave. In this case, the fragmentation into multiple  bow shocks could be
inhibited. In addition, there is mounting evidence that ambipolar diffusion is important in
protostellar outflows \citep{2002RMxAC..13...36D,2003MNRAS.339..524S,2004A&A...419..999G}.
Ion-neutral friction allows the magnetic field to efficiently
cushion the shock, spreading and combining the compression and cooling zones. This
would have two major effects. Firstly, molecules are not so effectively destroyed
until shock speeds approaching 50~km\,s$^{-1}$ are reached. Hence, the atomic emission
will be even more tightly distributed than here. Secondly, molecular emission is
generated from a wider range of shocks. This will tend to increase the length of
the wings of bow shocks and the flanks of cavities. Hence, MHD simulations which are directly
comparable to those presented in this series should provide more insight.

\section*{Acknowledgments}

MDS acknowledges INTAS support through grant 0-51-4838. This research was supported 
by the Particle Physics and Astronomy Research Council (PPARC) 
and the Northern Ireland Department of Culture, Arts \& Leisure.
The numerical calculations were executed on the Armagh SGI Origin 2000 computer (FORGE),
acquired through the PPARC JREI initiative with SGI participation, although
much of the analysis was performed on a desktop computer acquired through
COSMOGRID funding, which also funds AR.  As such, some of this work was carried out 
as part of the CosmoGrid project, funded under the Programme for Research in Third 
Level Institutions (PRTLI) administered by the Irish Higher Education Authority
under the National Development Plan and with partial support from the European
Regional Development Fund.

\end{document}